\documentclass[sigconf]{acmart}
\makeatletter                   
\def\mdseries@tt{m}             
\makeatother                    
\usepackage[plain]{fancyref}
\pdfoutput=1
\usepackage{color}
\usepackage{hyperref}           
\hypersetup{
    colorlinks=true,
    linkcolor=blue,
    filecolor=red,      
    urlcolor=magenta,
    breaklinks=true,            
}

\usepackage[utf8]{inputenc}
\usepackage{algorithm}
\usepackage{algpseudocode}

\usepackage[draft]{pdfpages}
\usepackage{svg}
\usepackage{subcaption}

\usepackage{xcolor}

\title{SMaLL: Software framework for portable Machine Learning Libraries}
\author{Upasana Sridhar}
\email{upasanas@cmu.edu}
\author{Nicholai Tukanov}
\email{ntukanov@cmu.edu}
\author{Elliott Binder}
\email{ebinder@cmu.edu}
\author{Tze Meng Low}
\email{lowts@cmu.edu}
\affiliation{%
  \institution{ECE, Carnegie Mellon University}
  \city{Pittsburgh}
  \state{PA}
  \country{USA}
  \postcode{15213}
}
\author{Scott McMillan}
\email{smcmillan@sei.cmu.edu}
\affiliation{%
  \institution{Software Engineering Institute, Carnegie Mellon University}
  \city{Pittsburgh}
  \state{PA}
  \country{USA}
  \postcode{15213}
}

\author{Martin D. Schatz}
\email{mschatz@fb.com}
\affiliation{%
  \institution{Meta}
    \country{USA}
}

\begin{document}
\begin{abstract}
   
Interest in deploying Deep Neural Network (DNN) inference on edge devices has resulted in an explosion of the number and types of hardware platforms to use. 
While the high-level programming interface, such as TensorFlow, can be readily ported across different devices, high-performance inference implementations rely on a good mapping of the high-level interface to the target hardware platform. Commonly, this mapping may use optimizing compilers to generate code at compile time or high-performance vendor libraries that have been specialized to the target platform. %

Both approaches rely on expert knowledge  to produce the mapping, which may be time-consuming and difficult to extend to new architectures.

In this work, we present a DNN library framework, SMaLL, that is easily extensible to new architectures. The framework uses a unified loop structure and shared, cache-friendly data format across all intermediate layers, eliminating the time and memory overheads incurred by data transformation between layers.
Layers are implemented by simply specifying the layer's dimensions and a \textit{kernel} -- the key computing operations of each layer.

The unified loop structure and kernel abstraction allows us to reuse code across layers and computing platforms. New architectures only require the 100s of lines in the kernel to be redesigned.
To show the benefits of our approach, we have developed software that supports a range of layer types and computing platforms, which is easily extensible for rapidly instantiating high performance DNN libraries.

We evaluate our software by instantiating networks from the TinyMLPerf benchmark suite on 5 ARM platforms and 1 x86 platform ( an AMD Zen 2).
Our framework shows end-to-end performance  that is comparable to or better than ML Frameworks such as TensorFlow, TVM  and LibTorch.

\end{abstract}
\maketitle
\section{Introduction}
    

The rapid development of the machine learning domain has led to
an explosion in the number of different computing platforms on
which deep learning models have been deployed. Today, there are
multiple GPUs, CPUs, and FPGAs that ship with specialized
hardware units (also known as accelerators) such as matrix engines~\cite{matrix_engines} and tensor cores~\cite{a100}
that speed up the computation of deep learning
layers. Specialized machine learning systems on the chip-scale are also rapidly being
introduced~\cite{ml_chips}. Last but not least, machine learning models are also
increasingly being deployed on embedded platforms that are
limited in both compute and memory resources.

Unlike the rapidly changing computing platforms, programming
interfaces have remain relatively stable with ML frameworks such
as TensorFlow~\cite{tensorflow} and Pytorch~\cite{pytorch} being the predominant
interfaces \emph{du jour}. This discrepancy in the rate of
evolution between the high level programming frameworks and the
underlying computing platform means that there is a need for a
way to quickly develop high performance, machine specific implementations of the high-level interfaces.

The two main approaches for mapping high level frameworks to
hardware specific back-ends are through the use of 1) custom high
performance libraries targeting the specific platform, or 2)
optimizing compilers that generate platform-specific codes at
compile time. Development of high performance back-ends (e.g.  oneDNN~\cite{onednn}, CMSIS-NN~\cite{cmsis-nn}) require
significant investment in time and manpower by expert developers
which makes it difficult for them to target the rapidly changing
platforms available. As such, many tinyML devices do not have
high performance back-ends that are specialized for the specific
embedded devices. Optimizing compilers also require significant amount of expertise to develop. However, the optimization heuristics in compilers are machine specific and need
to be tuned to avoid decisions that may not be suitable for specific devices (e.g. data reformatting that requires additional memory for tinyML devices).
Essentially, both methods are black-box
approaches which require highly-skilled experts with domain knowledge. Thus, it is difficult to extend to new architectures quickly. 

In this paper, we address this gap between high-level ML frameworks and platform specific code through the introduction of an open and portable framework that can be used as a back-end for the high-level ML frameworks. Specifically, the key observation behind our SMaLL framework is that many commonly encountered ML layers can be unified through a common abstract layer. This allows us to develop a single loop nest that can be specialized for various ML layers via a small set of parameters, and the provision of a specialized function for each layer. 
This common loop nest benefits from  the use of a common data layout across many different layers, and the isolation of performance/platform-specific code into small pieces of code (\emph{kernels}). Common data layouts across different layers eliminates the need for data reshape and packing that increases the memory footprint of the ML network; an overhead that is not conducive for deploying ML networks on edge devices. The identification of the required kernels reduces the effort of implementing entire high performance libraries to just the implementation of the kernels, which makes it easier for developers to port the framework to edge devices that often do not have high performance libraries as the computation back-end. Furthermore, the openness of the proposed framework allows others to develop custom kernels that target specific architectural features.
We show that despite the open and modular nature of our approach, our framework attains competitive performance against implementations that use different ML frameworks and libraries.
Furthermore, we demonstrate performance portability by using the same framework (with different custom kernels) on both tinyML, mobile and regular CPU devices.



\section{Background}

\subsection{Expert ML Libraries}
Many ML libraries such as oneDNN~\cite{onednn}, CMSIS-NN~\cite{cmsis-nn}, are high performance libraries that are developed by teams of expert developers over many years, and support numerous commonly-used ML layers.

However, many of these layers are implemented separately from each other. Very often, each layers has its own stand-alone loop nest, and the code may be replicated numerous times for different layers.  This results
in a large and extensive code-based that makes is laborious to port in a performant manner across different platforms. 

Moreover, 
 each library often
include multiple algorithms and implementations for each individual layer. These implementations may often have different data layout requirements, which means that data reshape routines are often required to perform data permutations between the different layout requirements for consecutive layers. A classic example is the need to perform the \texttt{im2col} transformation, either implicitly or explicitly~\cite{enrique_paper, im2col}, in order to leverage high performance matrix multiplication routines. These need to support different layouts and routines that transform between different layouts further increases the size of the code-based that needs to be supported.

Instead, the SMaLL framework recognizes that many of these layers can implemented using the same loop nest structure, and same data format. This reduces the complexity of the overall code-base and eliminates the need to support different data formats across consecutive layers. This isolation of the performance-critical code to the kernels also simplifies the task of the expert developer as only the performance-critical code needs to be updated across different platforms.

\subsection{Compiler Approaches}
Many state-of-the-art ML frameworks opt for using an optimizing compiler or Just-in-Time code generation to produce high-performance machine learning code. The goal with this approach is to rapidly support new architectures, independent of the availability of vendor-specific libraries. This method is used in TensorFlow ~\cite{tensorflow}, PyTorch~\cite{pytorch} and TVM~\cite{TVM} to varying degrees.

Often, this process starts with the generation of a compute graph of the neural network. For instance, TensorFlow uses the Accelerated Linear Algebra (XLA) ~\cite{XLA} optimizing compiler to perform high-level analysis on the compute graphs. The graph captures operations as nodes and input and output dependencies as edges. The graphs may also include the sizes of the tensors and their data types. The compilers then perform optimization passes on the compute graph. Such passes may rewrite the graph to include data-transformation required for high-performance implementations of a layer, or fuse nodes together if a fused implementation is available and deemed beneficial. TVM uses fusion extensively and also uses hardware-specific parameters to decide how to schedule the computation of the network -- whether to parallelize a given layer, how much to tile, etc.

A key optimization performed by almost all automated tools is to identify opportunities to perform data reshape of one layout to a different data layout that may be more suitable for a subsequent layer. An example of such a transformation is to hide the cost of packing from the original data to a format that can leverage the use of a high performance matrix-matrix multiplication routine (e.g. im2col). Our framework seeks to use common data layouts across all layers that eliminates the need for data repacking or reshaping. This is ideal for tinyML devices since these devices are limited in memory resources, and can ill afford to incur unnecessary memory overheads.

The compiler approaches may sometimes perform all of the lowering steps from the compute graph to the executable. XLA, PyTorch's optimizing compiler Glow ~\cite{glow}, and TVM have abstract backends that can be specialized to different hardware platforms. This allows them to use high performance vendor libraries when available and a default implementation if not. This design may cause a large variation in performance characteristics across platforms with different levels of vendor support.
SMaLL also has reference implementations of different layers, but the effort required to write specialized libraries is reduced to writing kernels.

\subsection{Portable Linear Algebra Library Instantiation}
Our SMaLL framework for instantiating ML libraries adopts the same approach as the 
BLAS-like Library Instantiation Software (BLIS), an open source framework~\cite{BLIS1} for quick instantiation of optimized Basic Linear Algebra Subprograms~\cite{BLAS1, BLAS2, BLAS3} libraries. Specifically, BLIS distilled the de-facto Goto algorithm for high performance matrix-matrix multiplication~\cite{goto} into 
 a single architecture-specific computation microkernel and two packing routines~\cite{BLIS2}. By providing custom implementations of the microkernel and/or packing routines, and specifying 5 architecture-specific parameters to the framework,
 an entire BLAS library can be generated for a specific architecture. 
 
 Within the BLIS framework, the 5 architecture-specific parameters are used to customize an intermediate data format that  tile the problem into sub-problems that fit into the different levels of the memory hierarchy~\cite{blis4} of the particular computing platform.  All other details of the matrices (e.g. the symmetry, values along the diagonal) and the handling of the edge cases are managed by the BLIS framework, which then frees the developer from having to implement code for the different cases.

 In a similar manner, our SMaLL framework identifies critical sections of code that need to be high performance, and the surrounding loops around the kernels are written in platform-agnostic manner to provide portability across different architecture. Our framework also utilizes a common layout that can be parameterized based on architecture features. This common layout is used across all supported layers to reduce the programming effort required to support different data layouts.

\section{Unified View of ML Layers}
The SMaLL framework is built around the key observation that many commonly used deep neural net (DNN) layers are stencil operations that compute each output element through the use of a sliding window over a set of input values. This observation suggests that different layers can be implemented using the same loop nest, which allows us to reduce the amount of code that needs to be written to support a new layer. In this section, we describe a unified view of different common layers seen in DNN workloads. 

\subsection{Abstract DNN Layer}
We begin our discussion with a description of an abstract DNN layer that can be specialized to concrete DNN layers. An abstract layer consists of  an input tensor ($\mathcal I$), an output tensor ($\mathcal O$), and an optional weight tensor ($\mathcal W$). The computation is represented by a mapping function $\mathcal F$ that reduces a subset of input elements from $\mathcal I$ (and optionally a corresponding subset of elements from $\mathcal W$) to a single output element in $\mathcal O$. The input elements are reduced using an operation specified in $\mathcal F$. 

\subsubsection{Input and Output Tensors}
The input and output tensors for our abstract layers are three dimensional, i.e. they have dimensions $\mathcal I_H$, $\mathcal I_W$, and $\mathcal I_C$ representing the height, width, and channels. Note that the output tensor dimensions are labeled similarly, with  $\mathcal O_H$, $\mathcal O_W$, and $\mathcal O_C$ representing the output height, width, and channels, respectively. Multiple inputs and/or output tensors can be supported with the addition of a fourth dimension, $\mathcal I_B$, corresponding to a batch of inputs. For ease of discussion, we will make the simplifying assumption that $\mathcal I_B=1$.

\subsubsection{Mapping from Input to Output Tensor}
The mapping function $\mathcal F$ specifies a window of input elements that must be reduced to produce each output element. The size of this window is described using the parameters $\mathcal F_H, \mathcal F_W$, and $\mathcal F_C$ to specify the number of elements being reduced in each dimension. Elements in the height and width of the output are computed by ``sliding'' the window across the input tensor according to strides in the height and width dimensions, denoted by $S_H$, and $S_W$. The number of element in the height and width of the output is given by
\begin{align*}
    \mathcal O_H & = \dfrac{\mathcal{I}_H - \mathcal F_H}{S_H} + 1 &
    \mathcal O_W & = \dfrac{\mathcal{I}_W - \mathcal F_W}{S_W} + 1    
\end{align*}
Similarly, elements in the channel dimension may be produced by sliding the window in the channel dimension according to the stride $S_C$. The number of channels produced by this sliding operation is given by the parameter $G$, the number of groups.
\begin{align*}
 G &= \dfrac{\mathcal{I}_C - F_C}{S_C} + 1
\end{align*}
The  reduction operation  specified in $\mathcal F$ may involve scaling elements in the input tensor by a set of learned values called weights or filters. This introduces the need for a weights tensor, described below. 

Further, the same window of the input may be scaled by a different set of weights. We use the parameter $K$ to denote the number of different filters that are applied to the same group of input elements to produce different elements in the channel dimension of the output. With these 7 parameters, the 3 reduction window parameters $\mathcal F_H, \mathcal F_W, \mathcal F_C$, their associated strides  $\mathcal F_H, \mathcal F_W, \mathcal F_C$ and finally the batch size of filters applied to the same window of the input $K$, we have a complete description of the mapping from input tensor to output tensor. 

Including the elements produced by reusing input elements and the elements produced by sliding the reduction window in the channel dimension, the total number of output channels is given by the product of the number of groups, $G$, and the number filters in 1 group, $K$.

 By having the reduction operation be a parameter of the mapping function $\mathcal F$, layers with the same window parameters $\mathcal F_H \times \mathcal F_W \times \mathcal F_C$ can be grouped together.
\begin{align*}
 \mathcal{O}_C &= K \times G
\end{align*}
\begin{figure*}
    \centering
    \includegraphics[width=0.8\textwidth]{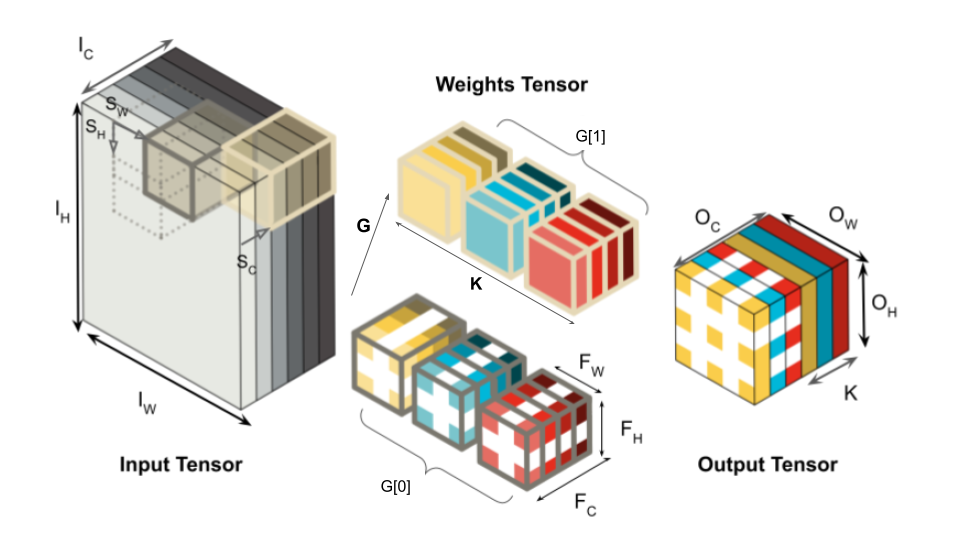}
    \vspace{-3mm}
    \caption{Visualization  of the abstract layer. The filter tensor $\mathcal{W}$ has $G= 2$ groups of filters with $K = 3$ filters in each group, producing a total of $O_C = 6$ output channels. Each filter $\mathcal{F}$ reduces $\mathcal{F}_W\times \mathcal{F}_H = 9$ input elements in  $ \mathcal{F}_C = 4$ input channels.}
    \label{fig:abstract_layer}
\end{figure*}





\subsubsection{Weight Tensor}
The weight tensor {$\mathcal W$} is optional and its dimensions depend on the reduction operation in $\mathcal F$. In the types of ML layers commonly used,  {$\mathcal W$} has
five  dimensions $\mathcal F_H \times \mathcal F_W \times \mathcal F_C \times K \times G$. The $\mathcal F_{H} \times \mathcal F_{W} \times \mathcal F_{C}$ elements describe the filter values elements in the reduction window of the input tensor are scaled by. $K$ accounts for the batch of filter elements that are applied to the same window. We bring special attention to $G$, a parameter derived from the window size and stride in the channel dimension. While elements produced by sliding the reduction window over the height and width of the input use the same filter, elements from different groups may be scaled by different filters. Intuitively, the channel dimension represents different attributes of the input, and  Thus, the weights tensor includes the $G$ dimension.

\subsubsection{Visualizing the mapping from Input to Output Tensor}
Pictorially, the different components of our abstract deep learning layer are shown in Figure~\ref{fig:abstract_layer}.
The input tensor is a 3 dimensional with a size $\mathcal{I}_H \times \mathcal{I}_W \times \mathcal{I}_C$. The weights tensor is grouped,in this case, into two groups, $ G_0$(central white cross) and $G_1$ (solid colors) each with $K = 3$ filters.The output tensor therefore has 6 channels, 3 with solid colors and 3 with the white cross pattern. Each filter also has a depth $F_C$ to reduce input channels. In the input, we see the stride between the groups $S_C$. The figure depicts a case where $S_C < F_C$, so there is some overlap in the input channels used across the two groups of filters. As mentioned above, by changing the reduction 

\subsection{Classes of layers}
Using the abstract DNN layer, we can specialize it into 5 main categories of layers that are often found in many deep learning models. Specifically, the layers can be differentiated by the dimensions of the reduction window, given by $\mathcal F$. The five specialized layers are as follows:





\begin{enumerate}
    \item\textit{Single Element Reduction}\\
    A single element reduction is also known as a point-wise operation. This class of layers reads one element from the input tensor and produces one output element. These layers can be characterized by the following filter parameters:
\[
\mathcal F_W = \mathcal F_H = \mathcal F_C = 1.
\] 

Activation layers using ReLU or Sigmoid are examples of this class of layers. Point-wise layers may involve a weight tensor whose elements are used as part of the computation of the layer. Addition
and batch normalization are examples of single element reduction layers. 

\item \textit{Single Channel Reduction}\\
These layers reduce elements from one input channel when computing a single output element. Specifically, the number of elements in the height and width of the input tensors are such that
\begin{align*}
\mathcal F_C = 1 \\
\mathcal F_H < \mathcal I_H, & \mbox{ and } \\
\mathcal F_W < \mathcal I_W.
\end{align*}
The size of the reduction in the input channel dimension is 1.
Depthwise Convolution and Pooling are examples of this type of reduction, where Depthwise Convolution involves the use of the weight tensor while Pooling does not.

A 2D max pooling layer reads in elements in a single input channel, and returns a single output element that is the maximum of all the $\mathcal F_H \times \mathcal F_W$ input elements.
A Depthwise convolution has $F_C = 1$ and a similar pattern of reducing $\mathcal F_H \times \mathcal F_W$ input elements in a single channel to produce a single output element. The reduction operation  requires a weight tensor with one set of weights for each input channel, i.e. $G = \mathcal I_C$. In most cases each input channel is used to produce a single output channel, so $ K = 1$. 


\item\textit{Partial Channel Reductions}\\
The size of the reduction in the channel dimension is partial along with the
height and width dimensions: 
\begin{align*}
1 < \mathcal F_W < \mathcal I_W, \\
1 < \mathcal F_H < \mathcal I_H, & \mbox{  and } \\
1 < \mathcal F_C < \mathcal I_C
\end{align*}
Group Convolutions are an example of this reduction. Input channels are divided into groups such that the $\mathcal F_C$ channels within a group are reused to produce $K$ output channels. 
Group Convolutions typically have a stride equal to the channel reduction $S_C = F_C$, so that the groups are completely disjoint. 

\item\textit{Full Channel Reductions}\\
These layers compute an output element using inputs elements from all input channels, i.e.
\[
\mathcal F_C = \mathcal I_C
\]
    
Examples of this sort of reduction are the traditional convolution and the $1 \times 1$ convolution found in neural nets such as VGG~\cite{VGGnet} and ResNet ~\cite{resnet}. The traditional 2D convolution reduces a small spatial window across all input channels to get 1 output pixel. Different filters are used with the same spatial window to get the different output channels. Given this, we see that the channel dimension of the filter must be equal to the total number of channels in the input, such that $\mathcal F_C = \mathcal I_C$. The number of output channels $\mathcal O_C$ is a hyper-parameter.

A $1 \times 1$ Convolution ~\cite{Inception} simply reduces all input channels corresponding to one pixel of the input to get each output element. Notice that this means all the sliding window parameters are the same as the traditional convolution, except that the spatial dimensions of the filter are now fixed and equal to 1 ($F_H = F_W = 1$). This configuration also means that the spatial dimensions are computed completely independent of each other and exhibit no overlap.

\item\textit{Full  Reductions}\\
This operation reduces all elements in the input tensor to produce one output element, i.e. $F_W = \mathcal{I}_W, F_H = \mathcal{I}_H$ and $F_C = \mathcal{I}_C$. A Fully Connected or Linear layer is an example of a full reduction. The input can be viewed as a flattened vector $1 \times (\mathcal{I}_H\cdot\mathcal{I}_W\cdot\mathcal{I}_C)$ that is multiplied with a $ (\mathcal{I}_H\cdot\mathcal{I}_W\cdot\mathcal{I}_C) \times K$ weights tensor to produce $K$ output elements.
\end{enumerate}
\begin{table}[h!]
\caption{Classification of layers based on number and location of input elements that are reduced to obtain a single output. Entries with $*$ indicate parameters that could vary depending on the type of layers.
The Partial Channel reduction has the most degrees freedom for mapping inputs to outputs. }
\label{tab:loop_classification}
\centering
 \begin{tabular}{lccccc}
        \hline \hline
        \bf Layer Class & $\mathcal F_H$ & $\mathcal F_W$ & $\mathcal F_C$ & $K$ & $G$ \\
        \hline
        Single Element Reduction & 1 & 1 & 1 & $1$ & 1 \\
        Single Channel Reduction & * & * & 1 &$ 1$ & $\mathcal I_C$\\
        Partial Channel Reduction & *&*&*&*&* \\
        Full Channel Reduction& * & * & $\mathcal I_C$& $\mathcal O_C$ & 1\\
        Full Reduction &  $\mathcal I_H$ & $\mathcal I_W$ &$\mathcal I_C$ & $\mathcal O_C$ & 1\\
        \hline \hline
    \end{tabular}
%
\end{table}
%
%
\subsection{Mapping specialized layers to actual DNN layers}
The 5 classes of layers presented above can be further specialized into actual DNN layers found in common ML networks. The essence of this mapping is the specification of the reduction function that forms the core of the operation that is performed given the elements from the input and weight tensors.
In Table \ref{tab:loop_mapping},  we provide the mapping for some representative DNN layers from each specialized layer class.

\begin{table*}[h!]
\caption{Representative layers for each specialized layer. The reduction function  captures the different computation across different layers. The stride dimensions are often input parameters that are determine by the ML networks. They define which input elements are mapped to output elements }
\label{tab:loop_mapping}

\centering
    \begin{tabular}{l c c cc c}
        \hline \hline
        \bf Layer Type & \bf Specialized Layer   & \bf $\mathcal S_H$ & $\mathcal S_W$ & $\mathcal S_C$ & \bf Reduction Function\\
        \hline
        Activation (ReLU) & Single Element &1 & 1 & 1 & Max\\
        Batch Normalization& &1 & 1 & 1 & FMA  \\
        Upsample& & * & * & 1 & Nearest Neighbor Interpolation \\
        \hline
        Depthwise Convolutions& Single Channel & * & * & 1 & FMA\\
        Max Pooling& & *&* & 1 & Max \\
        \hline
        Group Convolutions& Partial Channel & *&*&*& FMA \\
        \hline
        Convolution& Full Channel& *&*& 1 & FMA \\
        \hline
        Fully Connected& Full &1 &1 &1 & FMA \\       
        \hline \hline
    \end{tabular}
\end{table*}

\section{Framework for Implementing ML Layers}
The key observation behind the unification of many commonly encountered ML layers through the abstract layer is the suggestion that a single loop nest can be designed for all the different layer classes. Furthermore, different actual ML layers can be obtained when specialized reduction functions are provided as the computational kernel for each layer. We discuss how these insights can be instantiated in a common framework that facilitates performance and portable implementations across different architectures.

\subsection{Common High Performance Loop Nest}
We begin with a discussion on the design of a common high performance loop nest for the abstract ML layer.
Notice that in order to compute each output element, we first need to iterate over $\mathcal F_H \times \mathcal F_W \times \mathcal F_C$ input elements. In total there are $\mathcal O_H \times \mathcal O_W \times \mathcal O_C$ outputs. Hence, a minimum of 6 nested loops are required to compute an abstract layer.

It has been shown that these 6 nested loops can be reorganized such that a direct convolution can be implemented in a high performance manner that is competitive with a matrix-matrix multiplication based implementation without the need for additional memory for data packing~\cite{direct_conv}. At the crux of the high performance loop nest is the use of additional loops that block/tile the current loops such that they map better to the available hardware resources such as the number of registers. The value for the parameters in these additional loop nests can be derived from the specific architecture
architecture.

Using the same approach, we can define the loop nest for SMaLL as shown in 
Algorithm~\ref{alg:direct_conv}. As the direct convolution is a specialization of the abstract layer, a simple change to the high performance direct convolution loop structure is required to generalize it in order to target different layer classes. Much of the performance analysis remains the same, with the innermost loops iterating over independent output elements, irrespective of the type of reduction. Across all layers, the bounds of the innermost layer represent 

The key difference between the loop nest proposed by Zhang et. al.~\cite{direct_conv} and the algorithm in Algorithm~\ref{alg:direct_conv} is 
that the loop over elements in the output channels from Zhang et. al. is split into a loop over the group parameter $G$, and over the number of filters $K$. This loop nest allows us to compute Single Channel Reductions since this particular layer is required to iterate over the $G$ input channels in order to compute a single output element. 
The splitting of the output channel loop into loops over the dimensions $G$ and $K$ also means that the 
the output channels can now belong to the same group or be distributed across groups. This means that the original blocking parameter $C_b$ that partitions the output channels into channel blocks of $C_b$ channels is also split into a tiling parameter on channels within a group $K_b$ and another on the groups of channels $ G_b$. Similarly, the block $C_b$ input channels are derived from the product of a new tiling parameter $F_{Cb}$ on the channel dimension of the reduction and the group tiling parameter $G_b$.

\color{black}
\begin{algorithm}
\caption{Abstract layer loop nest}\label{alg:direct_conv}
\begin{algorithmic}[1]
\For{$g \leq G$ in step $G_{b}$ in parallel}
    \For {$k \leq K$ in step $K_{b}$ in parallel}\label{channel_block}
        \For {$i \leq F_C $ in step $F_{Cb}$}
        \color{black}
          \For {$j \leq O_{H}$}\label{height} 
          \color{black}
                \For {$l \leq O_{W}$ in step $O_{Wb}$} 
                    \State{}    \Comment{kernel}
                    \For {$x \leq F_{H}$} 
                        \For {$y \leq F_{W}$}
                            \For{$ii \leq F_{Cb}$}
                                \For {$ll \leq O_{wb}$} \label{pixels}
                                    \For {$ gg \leq G_{b}$}
                                        \For {$kk \leq K_{b}$}\label{channels}       
                                        \State{ReductionOp($\mathcal O, \mathcal W, \mathcal I,$ }
                                        \State{$\quad \quad \mathcal S_H, \mathcal S_W, \mathcal S_C$)}        
                                        \EndFor
                                    \EndFor
                                \EndFor
                            \EndFor
                        \EndFor
                    \EndFor
                \EndFor
            \EndFor
        \EndFor
    \EndFor
\EndFor     
\end{algorithmic}
\end{algorithm}

\subsection{Specialized Kernels for Different Layers}

Having a common loop structure allows us to iterate over all input, output and weight elements across all layer classes in a unified manner. However, to instantiate them into actual DNN layers, we need to identify and implement specialized reduction functions, which we will term \emph{kernels}, for each actual layer that needs to be instantiated. Furthermore, these kernels have to be implemented in a high performance/hardware-specific manner in order to attain an efficient implementation of the actual layer.

Notice from Algorithm~\ref{alg:direct_conv} that the computation (i.e. the reduction function) is performed in the innermost loop. Furthermore, the immediate loops surrounding the computation are loops around different output elements and/or input elements. Recall that each specialized layer only differs from each other in 5 different parameters; all of which are captured by lines 7-12 of Algorithm~\ref{alg:direct_conv}. This means that in order to specialize the abstract layer into the different classes, we only need to write code that replaces lines 7-12. We term this specialized code as the kernel for each specialized layer.

\subsubsection{Design of kernel}
The kernels are defined as the innermost loops of Algorithm~\ref{alg:direct_conv}. To obtain high-performance, these kernels are written in a platform-specific manner. By themselves, the kernels only perform a partial update of a small segment of overall output{\footnote{This is in contrast to say a GPU kernel that requires the entire computation to be placed on the GPU. Similarly, the TFLite delegate/kernel model requires entire subgraphs of comptutation to be placed on the GPU/accelerator.}}. Specifically, the core design of each kernel is based on the computation that needs to be performed to partially update $O_{wb} \times G_b \times K_b$ output elements. For each of the output elements that is being computed, $\mathcal F_H \times \mathcal F_W \times \mathcal F_{Cb}$ input elements (and potentially the same number of weights) are read. The loops around the kernels allow us to use a fixed size kernel and use it to target different workload shapes and sizes. Edge cases are handled by code surrounding the kernels.

The specific values of the parameters $O_{wb}$, $G_b$, $F_{Cb}$ and $K_b$ that define the kernel are often based on machine parameters such as the latency, throughput of specific instructions, number of available registers, and the number of elements that can be computed per Single Instruction Multiple Data (SIMD) instruction. 
In order to determine the values for specific architectures, we follow the approach taken by Zhang et. al~\cite{direct_conv} and Low et. al~\cite{BLIS4x}. Since each kernel computes $O_{wb} \times G_b \times K_b$ independent outputs, the parameters of the kernels are sized to fit within the number of available SIMD registers. This ensures that the computation can be performed without register spilling, which incurs additional data movement overhead. Furthermore, because the computation is performed on independent outputs, this means that there are multiple independent instructions within the innermost loops that will exploit instruction level parallelism that is available. This requirement on the number of registers is often sufficient to fill in the pipeline so that the latency of the instructions can be hidden.

Moreover, the combined dimension of $G_b \times K_b$  represents the blocking on the channel dimension of the output, $O_{cb}$. The product of these two dimensions is the same across different layer types and is often set to be a multiple of the SIMD length. The specific values of $G_b$ and $K_b$ depend on the possible values of $K$ and $G$. For instance, given that $K = 1$ for a depthwise convolution $G_b = O_{cb}$


 While the computation required for different layers may vary, we can simplify the design of the kernel into 3 separate phases. First, the memory that will hold the updated output elements must be initialized (LOAD phase). Then, the reduction operation must be performed, which may include loading the appropriate input and weight elements (COMPUTE phase). Finally, the updated elements must be stored back to the output tensor(STORE phase).
 
Each update to the output during the COMPUTE phase is described by the loops on lined 10-12 in Algorithm \ref{alg:direct_conv}. The loops on lines 7-9 update the same tile of the output, so we move the LOAD and STORE phases outside of these loops. Algorithm \ref{alg:fma_reduction_kernel} shows the structure of a kernel for convolution. 
\begin{algorithm}
\caption{Specification of a Convolution Kernel with highlighted platform-specific vectorized phases}\label{alg:fma_reduction_kernel}
\begin{algorithmic}[1]
\Function{Kernel\_Convolution}{$\mathcal O, \mathcal W, \mathcal I, \mathcal S_H, \mathcal S_W, \mathcal S_C$}
    \State{LOAD\_VECTOR\_TILE($\mathcal{O}, O_{Wb}, G_b , K_b$)} \Comment{Set up vectors with output tile}
    \For {$x \leq \mathcal F_{H}$} 
        \For {$y \leq \mathcal F_{W}$}
            \For{$ii \leq \mathcal F_{Cb}$}
                \State{ FMA\_VECTOR\_TILE($\mathcal{I}, \mathcal{W}$)}\Comment{Platform specific compute phase}
            \EndFor
        \EndFor
    \EndFor
    \State{STORE\_VECTOR\_TILE($O_{Wb}, G_b , K_b, \mathcal{O}$)} \Comment{Store partial update to output}   
\EndFunction
\end{algorithmic}
\end{algorithm}




\subsection{Performance Portability}
A key observation we made is that the implementations of these 3 phases can often be reused across different actual layers. For example, the LOAD phase loads data from main memory into SIMD/vector registers. This action is often repeated across different layers and is independent of the reduction function that is performed. As such, only a single high- performance platform specific LOAD phase need to be instantiated for the ML layers in the same layer class. Similar observations can be made for the STORE phase.

The recognition that many phases across different layers are essentially identical allows us to greatly reduce the amount of code that has to be written to instantiate all the layers supported by the SMaLL framework. Specifically,  this observation allow us to leverage the framework to rapidly instantiate a ML library by reusing code (through the use of templates and macros) such that a 10k LoC ML-library requires only on the order of 100s of lines to be re-written for the specific architecture.




\subsection{Common Data Format}
\begin{figure}
    \centering
    \includegraphics[width=0.5\textwidth]{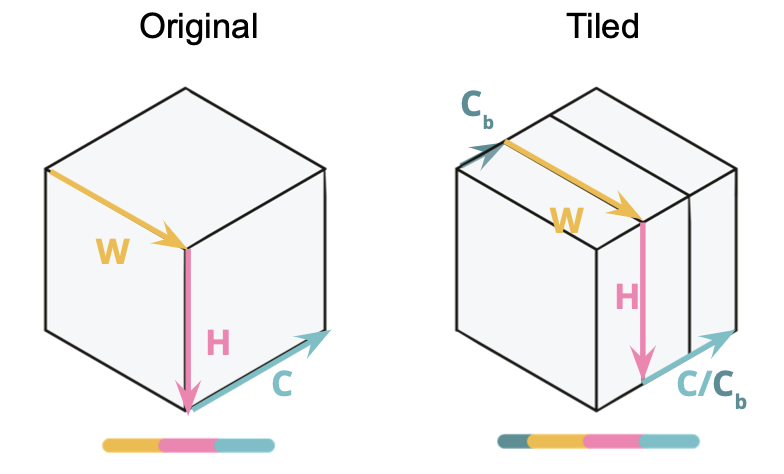}
    \caption{Common data layout for input and output across layers. The stripe at the bottom represents the dimensions from fastest to slowest.$C_b $ is a machine-specific parameter. All other are defined by the problem size  }
    \label{fig:data_format}
\end{figure}

Despite the slight difference in the loop nests in Algorithm~\ref{alg:direct_conv}, and the loop nest proposed by Zhang et. al.
~\cite{direct_conv}, we adopt the same data layouts proposed Zhang et. al. Specifically, two separate layouts were proposed in the prior work: one layout for the input/output tensors, and another for the weights tensor. We provide an overview of the two data layouts but refer the reader to Zhang et. al.~\cite{direct_conv}, for the rational behind the different layouts.

The data layout we adopted tiled the  channel dimension for the input and output tensors. The fastest dimension is the number of channels ($C_b$) and is a machine-specific parameter in Fig \ref{fig:data_format}. The next two fastest dimensions are the width then height dimensions. The slowest dimension is over the blocks of output channels. The layout of the output tensor is identical to the input tensor as the authors of the work made the observation that the input of a latter layer is usually the output of a prior layer.

The weight tensor is then tiled in both the input channel dimension and the output channel dimension to be compatible. The order of the dimensions in the data format matches the loop structure in Algorithm \ref{alg:direct_conv}.







\section{Results}\label{sec:Results}
\subsection{Experimental Setup}
Experiments were conducted on an array of edge devices as well as a desktop platform. The list of architectures on which custom SMaLL implementations were ran is summarized in Table \ref{tab:platforms_comparisons}. Furthermore, the comparisons with other ML libraries and/or frameworks are also described in the table.


\begin{table*}[!t]
\small
    \centering
    \caption{Summary of microarchitectures and platforms tested using high-performance SMaLL kernels, along with data type used during the comparison. A `-' indicates that the library/framework does not support the particular architecture. } 
    \begin{tabular}{p{20mm}|p{20mm}|p{15mm}|p{15mm}|p{15mm}|p{10mm}|p{15mm}|p{10mm}}
    \hline
    \hline
        \textbf{Platform} & \textbf{Micro-architecture} & \textbf{SMaLL} & \textbf{TensorFlow} & \textbf{TFLite} & \textbf{Libtorch} & \textbf{TVM/ uTVM} & \textbf{SNPE} \\ \hline
        Raspberry Pi 4B & ARM Cortex-A72 & float & float & float & - & float (tuned) & - \\ \hline
        Ryzen 3600x & x86 AMD Zen2 & float & float & float & float & Not completed & - \\ \hline
        Snapdragon 888 Kryo 680 (Silver) & ARM Cortex-A55 & float & - & - & - & - & float \\ \hline
        Snapdragon 888 Kryo 680 (Gold) & ARM Cortex-A78 & float & - & - & - & - & float \\ \hline
        Snapdragon 888 Kryo 680 (Prime) & ARM Cortex-X1 & float & - & - & - & - & float \\ \hline
        Arduino Nano 33 BLE & ARM Cortex-M4 & quantized uint8 & - & Failed & - & quantized uint8 (untuned) & - \\ 
    \hline \hline
    \end{tabular}
    
    \label{tab:platforms_comparisons}
\end{table*}


The Armv8 Raspberry Pi 4, a single board computer (SBC), the x86  AMD Zen 2, a consumer grade desktop CPU and the 3 platforms on the Kryo 680 CPU ~\cite{arm-cortex} on the Snapdragon 888 Mobile Soc ~\cite{Snapdragon}.

Our implementations are written in C++ and compiled using \texttt{gcc 9.4.0} on the AMD CPU and \texttt{gcc 10.2.1} for the Raspberry Pi. We use TensorFlow 2.10 as our baseline for the desktop CPU. TensorFlow, combined with optimization from XLA represent the state of the art for performance, so we also compare our library against the implementations in TensorFlow executed eagerly with Just-in-Time compilation from XLA which uses the AVX2 and FMA instruction sets~\cite{intel_extensions}. We refer to this combination simply as TensorFlow for the rest of this section. LibTorch (v1.8) is a C++ interface to PyTorch, which we use to compare our SMaLL implementation. SIMD instruction sets. 

For the Raspberry Pi, we compare against TensorFlow and the lightweight \texttt{tflite\_runtime} package as our baseline. To the best of our knowledge, LibTorch is not supported on ARM platforms, so we do not report a comparison to it.  We also present a comparison to a tuned TVM ~\cite{TVM, autoTVM}inference module on the Raspberry Pi.
We also include TFLite ~\cite{tflite} as part of our comparisons. TFLite uses a pre-compiled compute graph of the network and is able to perform advanced optimizations such as layer fusion on this graph.

For the different CPUs that make up the Snapdragon 888 Kryo 680, we compiled SMaLL implementations on a Linux device using the GNU C++ compiler for ARM platforms Version 9.4.0 (\texttt{aarch64- linux-gnu-g++) and Ubuntu 9.4.0-1ubuntu1~20.04.1}. As a comparison point we use the Snapdragon Neural Processing Engine (SNPE) version 2.5.0.4052~\cite{SNPE}. Unlike the AMD CPU and the Raspberry Pi, these implementations were written on an external device and cross-compiled for the specific target.  

Additionally, we show that the framework is easily adaptable to inference with different data types by presenting results for quantized inference on an Arduino Nano 33 BLE, a Cortex M4 device running at 64 MHz. We run end-to-end inference with the same tiny models as above, with the inference being quantized to integer (unsigned 8 bit). The models, in SMaLL, were compiled using the Arduino IDE version 2.0.4 ~\cite{arduino-ide}. TVM used the Arduino CLI version 0.30.0 ~\cite{arduino-cli} to transfer models to the Nano.

\subsubsection{MLPerf:Tiny Inference Models}
MLPerf:Tiny~\cite{mlcommons_tiny} is a benchmark suite for TinyML systems. It features 4 models representative of the types of ML tasks that might be offloaded to embedded edge devices. All models except Deep-Autoencoder start with a traditional convolution layer. The subsequent layers in the network often have a recurring pattern of layers that are stacked together to form the network. 
Of the 4 models, 2 share a common building block -- the Depthwise Separable Convolution block ~\cite{ds-conv}. These are MobileNet~\cite{mobileNetv1} and DS-CNN~\cite{DS-CNN}, which use depthwise convolutions and $1\times 1$ convolutions. 
ResNet~\cite{resnet} uses the Residual Block, where the input to the block is accumulated with the output of the block. Each block is a stack of two $3 \times 3$ convolution layers. If the size of the output is different from the input, there is an additional $1 \times 1$ convolution on the input to match sizes before the accumulation with the output.Finally, the Deep-Autoencoder uses a stack of fully connected layers.
The models are summarised in Table \ref{tab:tinyml_models}.

\begin{table}
    \centering
    \begin{tabular}{|p{12mm}|p{14mm}|p{12mm}|l|}
    
    \hline
        \textbf{Model} & \textbf{Use Case} & \textbf{Input to Model} & \textbf{\ Params} \\ \hline
        MobileNet & Visual Wake Words (VWW) & (96, 96, 3) & 3201472 \\ \hline
        ResNet & Image Classification (IC) & (32, 32, 3) & 77744 \\ \hline
        DS-CNN & Keyword Spotting (KS) & (25, 5, 3) & 20288 \\ \hline
        Deep Auto-encoder & Anomaly Detection (AD) & (1, 1, 128) & 133120 \\ \hline
    \end{tabular}
\caption{A summary of the 4 models we use to compare end-to-end inference performance. The parameters column lists the total number of weights across all the layers \label{tab:tinyml_models}}
\end{table}

\subsection{Overhead of 
using Unified Loop Structure against TensorFlow}\label{subsec:layer_comparison}
We demonstrate that our proposed unified loop structure can be specialized into many of the commonly-used ML layers with minimal loop overheads by comparing 5 different layers against the equivalent implementations from TensorFlow: 1) ReLU and 2) Max Pool, which use the same Max kernel; 3) the standard 3x3 Convolution, 4) the special case of a 1x1 Convolution,  and 5) a Depthwise (Dwise) convolution. This set of 5 layers covers 3 out of the 5 types of layers described in Table~\ref{tab:loop_classification}. Moreover, this selection of layers are the layers used by the models in the MLPerf:Tiny benchmark~\cite{mlcommons_tiny}.

With TensorFlow, we use the implementations of common neural network layers available within the \texttt{keras} submodule\cite{keras}. We also run each layer through the Accelerated Linear Algebra (XLA) domain-specific compiler\cite{XLA} to access platform specific implementations of each layer.



With SMaLL, we evaluate two implementations. One implementation uses kernels written specifically for the Raspberry Pi's ARMv8 microarchitecture using explicit NEON vector intrinsics. In addition to our hardware-specific kernels, we evaluate the performance achieved by our reference kernels written purely in C++.
The reference kernels use the same parameters, $O_{wb} \times K_b \times G_b$ as the hardware-specific kernels, but the kernel is pure C++ code and does not explicitly use NEON intrinsics. The performance difference between these two implementations highlights the impact of specializing the implementation of the kernels by explicitly using hardware-specific instructions.

\subsubsection{Choice of layer types and sizes}
We choose 2 input tensor sizes from MobileNet as representative problems for each layer type. Both sizes compute the same total number of outputs but the elements are distributed across the height, width and channel dimensions of the output differently. The first problem size, a $48 \times 48 \times 16$ ($\textit{height} \times \textit{width} \times \textit{channels}$) input presents a case where the spatial dimensions (height and width) are larger and the channel dimension is smaller. The second problem size $6 \times 6 \times 512$ instead has a larger channel dimension with smaller spatial dimensions. This exercises how the frameworks handle variability in dimension sizes that is common in DNNs.


\subsubsection{Layer-by-layer performance comparison}

\begin{figure*}
    \centering
    \includegraphics[width=0.98\textwidth]{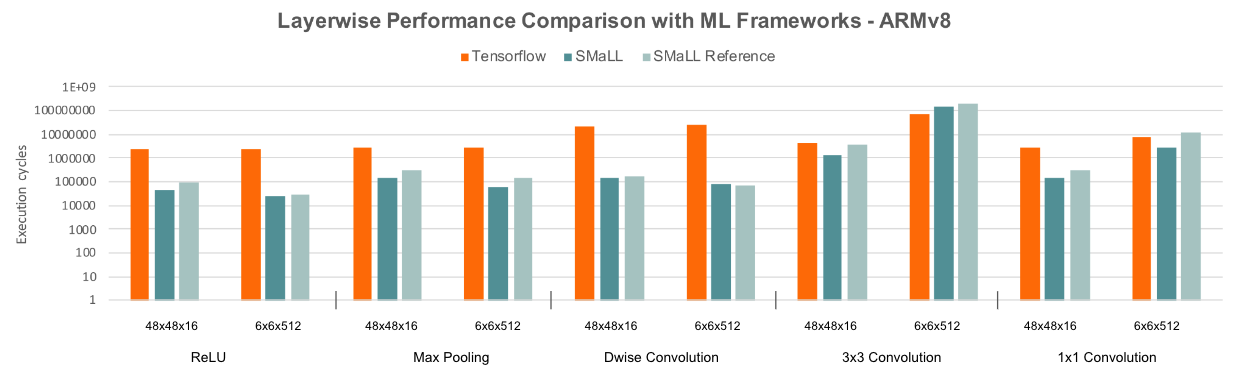}
    \caption{Comparing execution cycles of different layers using the unified loop structure implementations (SMaLL) against the TensorFlow implementations. 
    The vertical axis is log scale, lower is better. SMaLL, with hardware specific kernels , performs comparably on compute-intensive $3\times 3$ convolution layers and significantly better on all other layer types.  }
    \label{fig:layerwise comparison}
\end{figure*}

Figure \ref{fig:layerwise comparison} shows the minimum execution cycles over 100 trials for each framework on a Raspberry Pi 4 Model B.
We see that most layers are much faster using SMaLL with hardware specific kernels. It is also interesting to note that the reference implementation alone is faster than TensorFlow in many cases. However, the expert written kernels produce an additional 2 times speedup on average. This suggests that many of the individual layers in Tensorflow are not implemented in a high performance manner. This could be the result of having a very large code-base that makes it time-consuming for the expert developer to address all layers in a high performance manner. The observation that the large size $3 \times 3 $ convolution is the only Tensorflow implementation that out-performs SMaLL implementation suggest that this is the case that had been optimized.

Another possible reason for the SMaLL layers being faster is that the layout used in the SMaLL framework makes it easy to introduce kernels written using SIMD instructions. The layout in SMaLL ensures that the input data have good spatial locality, while also ensuring that the outputs are contiguous. Hence, even without SIMD instructions, the improvement in locality would reduce the time spent waiting for data to be brought into the processor.

As Tensorflow was compiled with XLA, the final implementation could have been composed and/or optimized from various different source implementations. When combined with an optimizing compiler, it is difficult to retroactively find the particular configurations being used at inference time. In contrast, since SMaLL is designed for an abstract layer, there is only a single code path for all problem sizes. Furthermore, 
every layer benefits from the performance decisions made at the abstract level. 

The depthwise convolution shows the most speedup from using SMaLL, with 135x and 311x speedup respectively. In XLA, the depthwise convolution is cast into a traditional convolution resulting in a significant increase in the amount of computation required. This could be a result of not having a specialized implementation for depth-wise convolution in XLA's list of optimizations. In contrast, SMaLL's implementation is a specific implementation of depthwise convolution that was instantiated using the abstract layer. This demonstrates the ease at which new layers can be supported using a common data layout and loop structure. More importantly, this shows that implementations with specialized filter sizes can be implemented in a high performance manner.

{\color{blue}{

}}


\subsection{Memory Usage Comparison}
We compared the memory utilization the SMaLL implementation against the implementation with TensorFlow Lite.  Table~\ref{tab:memory} shows the amount of memory allocated for the TinyML models in SMaLL, and the peak memory footprint of TFlite implementations reported by the TFLite Model Benchmark Tool~\cite{tf-lite-bench}. This metric is computed by running 100 inferences with a given model, performing a memory footprint check every 10ms and reporting the maximum observed memory footprint. Note that the tool reports that the number is approximate as the tool itself introduces some overhead.

The memory allocated by a SMaLL implementation is a function of the parameters required by the model (from Table \ref{tab:tinyml_models}) and the size of the intermediate buffers required to hold outputs of hidden layers. We recognize that output buffers can be shared across layers and only allocate space for the largest combination of input and output buffers required. For the floating point models, the memory requirement in bytes is the sum of the number of parameters and intermediate buffers multiplied by 4. For quantized models, even though the input and output values are uint8, accumulations to the partial result are in int32. Therefore, the memory required, in bytes, by the quantized models is the number of parameters, summed with the size of the intermediates, and 4 times the product of the largest height and width of any intermediate.

\begin{table}[tbh]
    \centering
        \caption{Comparison of memory used in SMaLL and TFLite. Memory usage is reported in MegaBytes(MB). Compared to the overall peak memory footprint of TFLite, SMaLL uses less memory}
    \begin{tabular}{l|lc|lc}
    \hline
    \hline
        \textbf{Model} & \multicolumn{2}{c}{\textbf{Floating Point}} & \multicolumn{2}{|c}{\textbf{Quantized uint8}} \\ \hline
           & SMaLL & Tflite & Small & Tflite \\ 
        autoencoder & 0.509 & 1.660 & 0.128 & 1.586 \\ 
        dscnn & 0.144 & 1.602 & 0.065 & 1.742 \\ 
        resnet & 0.433 & 4.867 & 0.202 & 4.055 \\ 
        mobilenet & 13.162 & 19.258 & 3.361 & 7.254 \\ \hline\hline
    \end{tabular}
    \label{tab:memory}
\end{table}

In all cases, the amount of memory incurred by SMaLL implementations is lower than their TFlite counter-part. This is expected since the SMaLL framework is designed around a single data layout format. This means that the SMaLL framework implementation does not require additional memory for packing data into different data formats. Each layer can directly operate on the output of the previous layer without any data reshapes. In comparison to a framework that might use temporary buffers to pack data into  specialized formats, SMaLL is expected to use less memory during inference time.

\subsection{End-to-End DNN Inference}\label{subsec:end-to-end}
\begin{figure}
    \centering
       \includegraphics[width=0.48\textwidth]{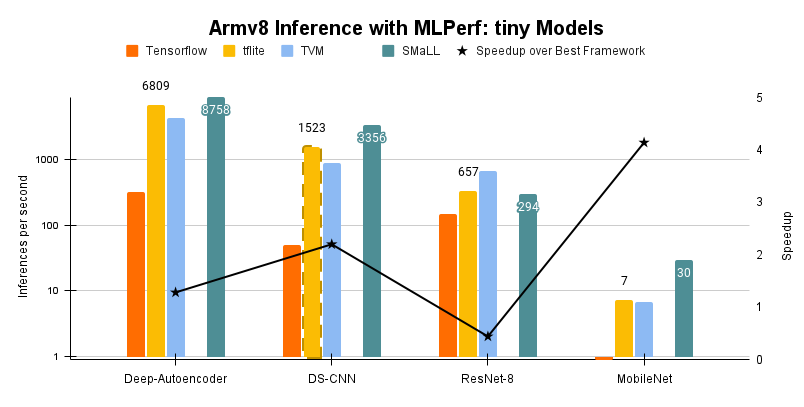}
    \includegraphics[width=0.48\textwidth]{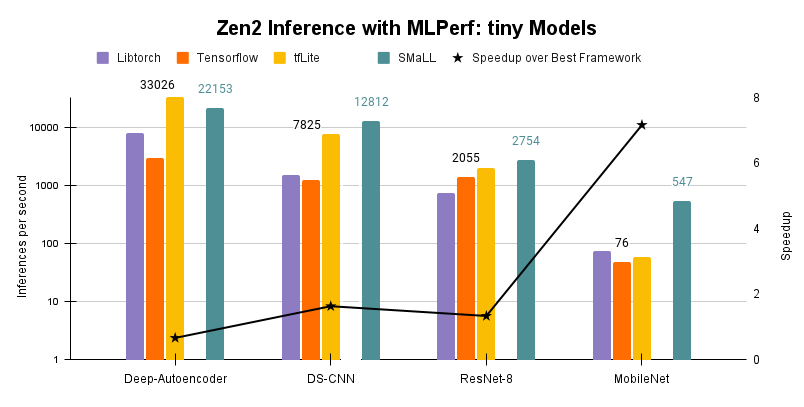}
    \caption{
    (Left) SMaLL gets comparable or higher framerate than TensorFlow TFLite and TVM. The dotted bar represents a modified network for TFLite, as the original network was unsupported. TVM attains the best performance on ResNet-8, which is mainly $3 \times 3 $ convolution. 
    (Right) State-of-the-art libraries has comparable performance with SMaLL but in many cases, SMaLL has the highest framerate. }
    \label{fig:tinymlperf_results}.  
    \end{figure}

\begin{figure}
    \centering
       \includegraphics[width=.32\textwidth]{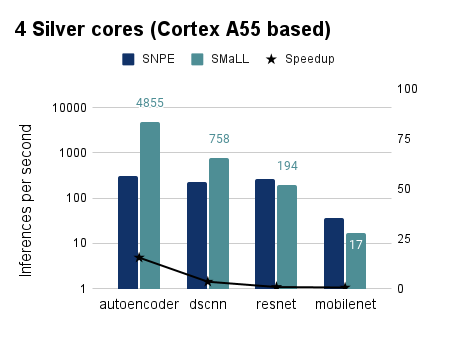}
        \includegraphics[width=.33\textwidth]{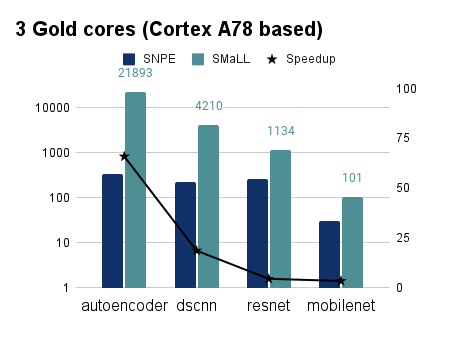}   
        \includegraphics[width=.31\textwidth]{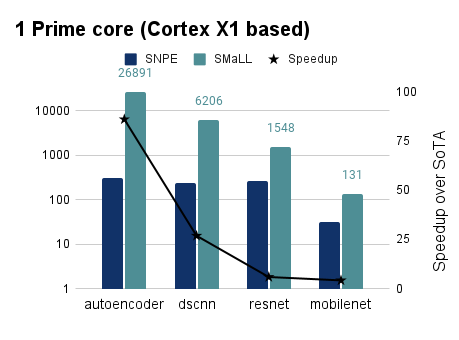}
    \caption{SMaLL outperforms SNPE on all 3 types of CPUs on the Kryo 680. SNPE performance is almost identical across all cores on the Kryo 680. 
    (Left) SMaLL is slower on 2 out of 4 models compared to the SNPE. This could be due to this processor being in-order. 
    (Center) Small attained between 3-65 times speedup on the Gold Cores.
    (Right) The comparison to SNPE shows an improvement of 4-85 times over SNPE on the Prime core.}
     \label{fig:MobileSoc-tinymlperf}
\end{figure}

We present the 4 MLPerf:Tiny Inference benchmarks on 6 platforms. The bars in Figures \ref{fig:tinymlperf_results} to \ref{fig:MobileSoc-tinymlperf} present the number of frames per second (FPS) achieved with each framework.

On the Raspberry Pi, in Figure \ref{fig:tinymlperf_results} (Left), we see the most speedup for the models that use the depthwise convolution -- 2x for the modified DS-CNN and 4x for the larger Mobilenet. As seen in our layer-by-layer comparison, SMaLL's depthwise convolution is much faster than TensorFlow's, so this result is expected. Autoencoder is made of fully connected layers. In SMaLL, these look very similar to a $1 \times 1$ convolution, which performs slightly faster, so we see a 28\% improvement over TFLite. ResNet has the most $3 \times 3$ convolutions of all the models, for which TensorFlow's implementation can be up to 2x faster than SMaLL's. We see that the end-to-end performance is slower for SMaLL, at 44\% of the framerate of the tuned TVM implementation. This may be due to the current parallelization scheme in SMaLL, where we only use 1 thread for inference. 



Note that, on the Raspberry Pi the rectangular filter ($3 \times 1$) for the first layer used in DS-CNN does not seem to be supported in TFLite, the outputs are always 0. The TFLite bar with dotted borders in the group for DS-CNN shows the framerate acheived for a model similar to DS-CNN where the rectangular filter is replaced with a square one ($3  \times 3$).  Also note that TFLite tends to be an order of magnitude faster than using TensorFlow. This difference may come from the lower execution time overheads from using the lightweight runtime system combined with the graph-level optimizations.

On the Zen 2, shown in Figure \ref{fig:tinymlperf_results}(Right), we see a similar trend of TFLite being faster than TensorFlow. Additionally it tends to be faster than LibTorch as well, with the exception of MobileNet. The rectangular filters appear to be supported for TFLite on the Zen2. 
SMaLL maintains good performance across the board, with the highest framerate for 3 out of the 4 models. The biggest speedups are once again for models with depthwise convolutions. 7.1x for MobileNet (over LibTorch) and 1.67x for DS-CNN. ResNet is 40\% faster than TFLite using SMaLL. The autoencoder model is 33\% slower than TFLite. This may be due to our simple parallization scheme that is unable to utilize the 6 threads we have on the Zen 2. 

On the mobile CPU, we use SNPE as a baseline. SNPE accepts pre-trained models in a model exchange format that it then optimizes and compiles into an executable(.dlc) for the Qualcomm board. We then use \texttt{snpe-parallel-run} to run each model, specifiying the CPU as the target using \texttt{-runtime-order cpu]}. 

Figure \ref{fig:MobileSoc-tinymlperf} compares the framerate achieved by SMaLL with that of SNPE for each of the 3 available CPUs. 
On the silver cores, (Figure\ref{fig:MobileSoc-tinymlperf} (Left)) SMaLL outperforms SNPE on autoencoder as well as DS-CNN. The larger models ResNet and MobileNet are faster using SNPE.  As we move from the lower throughput cores to the higher throughput cores, the performance gap widens, with SMaLL outperforming SNPE on all models. On the gold cores, autoencoder achieves a 65x higher framerate using SMaLL, mobilenet has a more modest 3x speedup. Finally, on the prime cores, SMaLL is 85x faster and mobilenet is 4 faster.

An analysis of layer-wise performance, suggests that the computation kernels for the Cortex A55, an in-order core, are not acheiving close to the peak throughput, while they are closer for the Cortex A78 and Cortex X1. This suggests a redesign of the kernels for the A55 might close the gap between SMaLL and SNPE.
Specifically, because the processor is in-order, a careful scheduling of the instructions is required to achieve better software pipelining~\cite{monica_lam} to hide the latency. While this could theoretically be performed by the compiler, our experience has been that a manual pipelining of the kernel is often beneficial.

Across all 5 platforms, we see that SMaLL consistently gets better performance for almost all models except ResNet on the Raspberry Pi where we are at 44\% of the FPS of the fastest framework with a single thread and the Deep Autoencoder on the Zen 2 where we are at 67\% of the FPS of TFLite. Further, SMaLL achieves good performance even when the model architecture has considerable layer diversity, a benefit that can be attributed to the universal design of the abstract layer.

\subsection{Scalability of Performance}\label{subsec:scalability}
\begin{figure}
    \centering
    \begin{subfigure}[b]{0.48\textwidth}   
        \centering 
        \includegraphics[width=\textwidth]{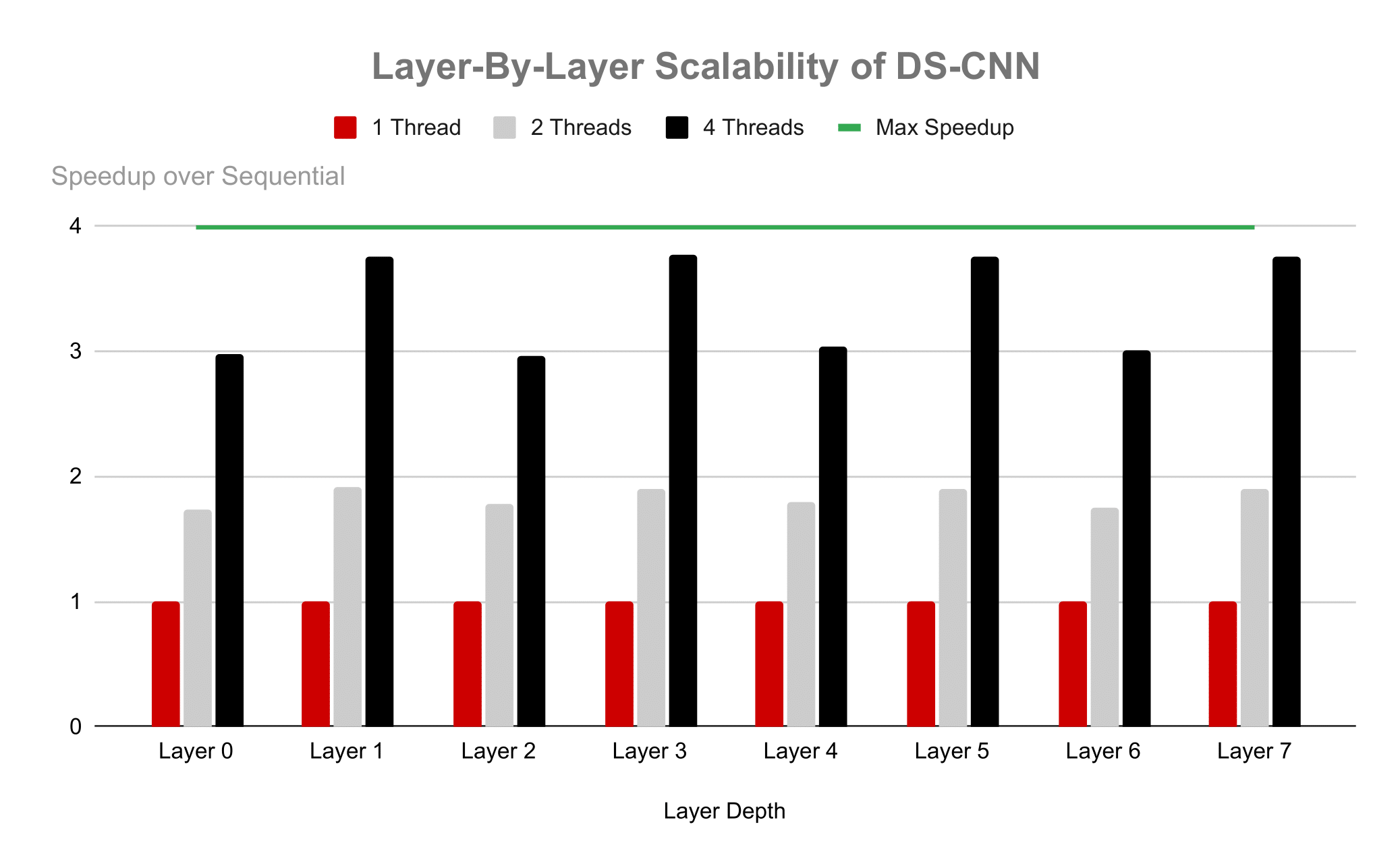}
        \caption[]%
        {{The  parallelism available across all the layers in DS-CNN is the same. }}    
        \label{fig:dscnn-parallel}
    \end{subfigure}
    \hfill
    \begin{subfigure}[b]{0.48\textwidth}   
        \centering 
       \includegraphics[width=\textwidth]{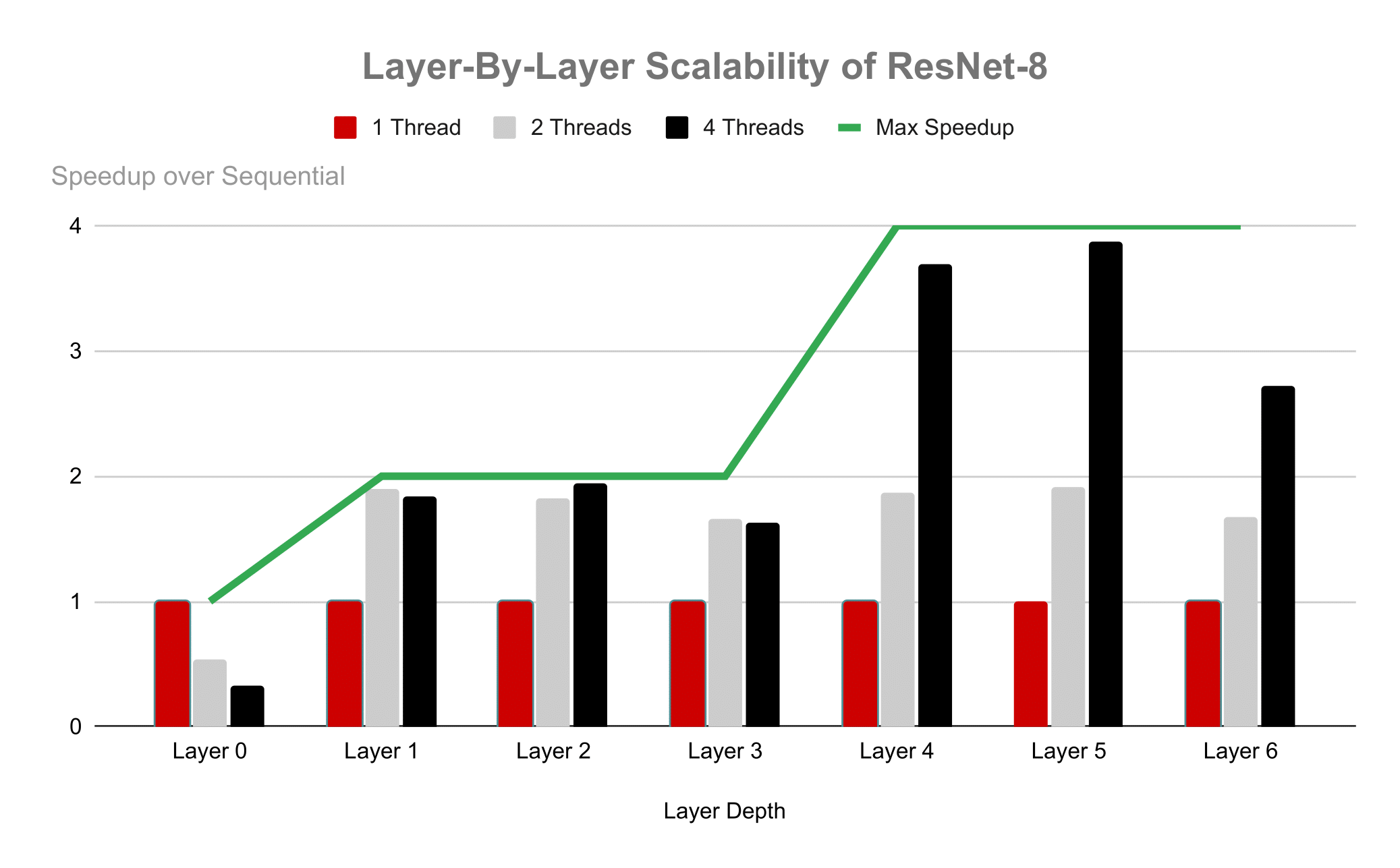}
        \caption[]%
        {{In Resnet, different layers have different amounts of parallelism available to SMaLL}}    
        \label{fig:resnet-parallel}
    \end{subfigure}
    \caption {{Study of the scalability of SMaLL to the 4 cores available on a Raspberry Pi. The extent to which the available parallelism can be leveraged depends on the dimensions of the output of each layer.}
        \label{fig:scalability}}
\end{figure}
While the focus of this work is not on parallelizing inference optimally, edge devices do tend to offer a small amount of multi-core parallelism, so we show how SMaLL can leverage this parallelism. The number of cores varies on our evaluated devices: the Zen 2 has 6 cores, the Raspberry Pi has 4 cores, the Kryo 680 CPU has 8 cores total, with 4 gold, 3 silver and 1 prime, finally  the Arduino Nano has 1 core.

We use the Raspberry Pi to illustrate scalability for layer sizes found in ResNet~\cite{resnet} and DS-CNN~\cite{DS-CNN} (Depthwise Separable Convolutional Neural Network). Further, both networks have a similar number of layers, and similar problem sizes.  DS-CNN uses  $3 \times 3$ depthwise convolutions (even-numbered layers in Figure \ref{fig:dscnn-parallel}) and $1 \times 1$ convolutions (odd-numbered layers in Figure \ref{fig:dscnn-parallel}). ResNet is largely comprised of $3 \times 3$ convolutions with a $1 \times 1$ convolution for Layer 3 and 6.

We parallelize the (blocked) channel dimension of the output tensor. The speedup we can expect from parallelism is limited by the size of this dimension. Figure \ref{fig:scalability} illustrates this limit with the 'Max Speedup' line, which assumes linear scaling of the single-threaded performance. For instance, in DS-CNN (Figure \ref{fig:dscnn-parallel}) all of the layers have 64 channels. With a block size of 16, all 4 cores on the Raspberry Pi can be used to parallelize the problem, resulting in the max speedup being 4x the single-threaded execution time. On the other hand, ResNet starts with a small number (16) channels, which is not parallelized. Layers 1-3 have 32 channels and therefore a max speedup of 2x. Layers 4-6 have 64 channels and can have a maximum speedup of 4x.
We choose the convolution-like layers to show scalability as these make up the majority of end-to-end inference time. However, we use the same parallelism scheme across all layers as a result of the unified loops.

In DS-CNN, all layers benefit from using 4 threads. In Figure \ref{fig:dscnn-parallel} the $1 \times 1$ convolutions are within 7\% of the maximum expected speedup. The Depthwise convolution layers scale at a slower rate, with a speedup of 3x on average with 4 threads. This may be because these layers have less overall computation that can amortize the cost of parallelism.

For the ResNet layers with expected speedup less than 4, we see that trying to use more threads than necessary can be severely detrimental. The $3 \times 3$ convolution layers scale to within 6\% of the maximum speedup irrespective of the problem size. The $1 \times 1$ convolution scales slower, achieving 75\% of the maximum speedup on average. This may be because the $1 \times 1$ performs less work than the $3 \times 3$ convolution per output element. Further, the layer stride is 2, leading to lower reuse of the input elements and therefore, a need to load inputs at a faster rate from lower levels of memory to compute the same number of output elements. As with the depthwise convolutions in DS-CNN, the cost of parallelism impedes the benefit from it.

We see that the amount of parallelism to be used for a given layer depends on the dimensions of the output tensor. Further, over-provisioning threads can be detrimental to layer performance. This suggests some trade-offs on the number of threads to use in the end-to-end case. We will see the results of applying these trade-offs in subsection \ref{subsec:end-to-end}.

\subsubsection{Parallelism for Entire Models}
In \ref{subsec:scalability} we showed that each layer has an available amount of parallelism that limits the number of threads used to parallelize it. Let us call this bound  $p_{avail}$. We also see that different layers within a single network benefit from different amounts of parallelism. In this subsection, we study how this diversity in available parallelism affects how much the end-to-end network must be parallelized. To do this, we set a ceiling $p_{max}$ for the amount of parallelism available to one inference. Each layer is parallelized over a number of threads $p$ given by $p = min(p_{avail}, p_{max})$.
\begin{figure}[h!]
    \centering
    \includegraphics[width=0.5\textwidth]{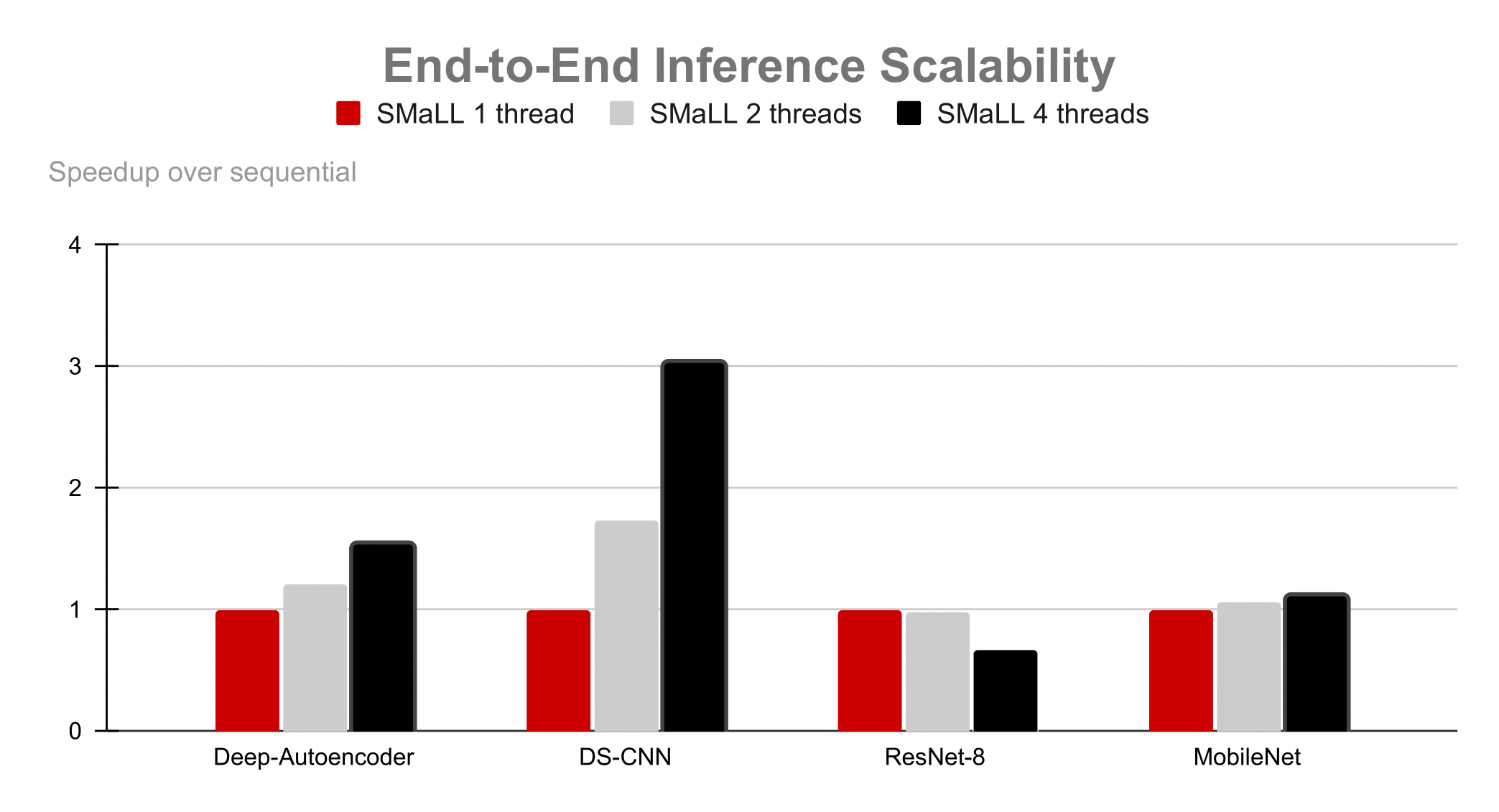}
    \caption{Speedup from using multiple threads for end-to-end inference. The read bar is single threaded. The varied amounts of parallelism in ResNet result in single threaded inference being the fastest. All other models benefit from using all 4 threads.}
    \label{fig:model_scalabilityl}
\end{figure}
Figure \ref{fig:model_scalabilityl} shows the minimum inference time over 100 runs with different parallelism ceilings $p_{max} = 1, 2, 4$. The fastest configuration is highlighted with a black outline. As Section \ref{subsec:scalability} would lead us to expect, DS-CNN performs the best with 4 threads due to all the threads having the same number of channels across layers. Autoencoder has a similar structure where 7 out of 8 layers have the same number of channels, 128. Mobilenet has more variance in channel size, from 32 to 1024. However all except the first layer has a $p_{avail} \geq 4$. So Mobilenet also performs the best when parallelized on 4 threads. The outlier is ResNet, where even though 6 out of 7 layers benefit from different amounts of parallelism, the best end-to-end performance is with a single thread. This may be because when each layer has a different $p_{avail}$, there is an imbalance in workload across threads.

\subsection{Quantized Inference}

\begin{figure}
    \centering
        \centering 
        \includegraphics[width=0.5\textwidth]{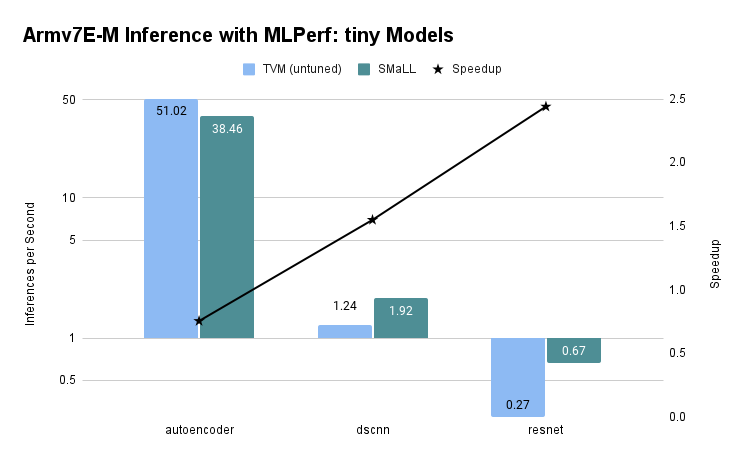}
        \caption[]%
        {{For the Armv7E-M platform, SMaLL gets comparable or higher framerate than TVM } \label{fig:pico-quantized-inference}}    
\end{figure}

The focus of the MLPerf:tiny Benchmark is small, resource-constrained devices. In this subsection, we show that SMaLL is readily portable to micro-controller platforms that use integer based inference.
We show experimental results on an Arduino Nano 33 BLE board which carries an Arm Cortex-M4 32-bit processor 
operating at 64MHz and 256KB of SRAM.
We compare SMaLL's performance of quantized networks against those produced by TVM v0.10.0. 
Note that Figure \ref{fig:pico-quantized-inference} does not feature Mobilenet. This is due to the memory requirement exceeding the 256K limit of the Nano. While the Nano does have a floating point unit (FPU) the memory requirements of the floating point models often exceed the memory limit. 

The performance achieved by small is comparable to TVM in all cases, with autoencoder being at 75\% of the TVM framerate and mobilenet being 2.4x faster. 

We also tried to provide a comparison with TFLite, but uploading the quantized models through the Arduino IDE caused the board to become unresponsive. This may be related to the overall peak memory consumption of TFLite, which, as reported in Table \ref{tab:memory} exceeds the 256KB SRAM of the Nano.

\section{Conclusion \& Future Directions}
In this work, we present the SMaLL framework, a open and portable framework for rapidly instantiating common encountered DNN layers for different architectures. At the core of the SMaLL framework, we leverage the insight that many ML layers can be unified using a single abstract layer. This abstract layer can be implemented through a single loop nest and the use of common data layouts, which minimizes the code that has to be written for each new layer, and for different architecture. The use of common layout across layers also eliminates the need for memory for data layout transforms, which facilitates the deployment of larger ML networks on memory-constraint devices.

We show that despite the modularity and high level of code reuse, the implementations provided through the SMaLL framework provided performance that is competitive with state-of-the-art machine learning frameworks, both on individual layers as well as on end-to-end inference with real models. Further, we showed current machine learning frameworks show very different performance characteristics on different architectures, this may be because they rely on vendor-optimized libraries. On the other hand, our framework allows the rapid porting of implementations to different architectures, thereby democratizing access to advanced machine learning techniques across devices.

While we have shown performance for 6 different architectures, this list is not exhaustive. Our approach extends easily to different platforms as demonstrated by successes in other numerical domains~\cite{BLIS1}. We are actively looking into adding specialized kernels for a wider range of edge computing platforms  such as the Nvidia Jetson ~\cite{jetson} and Qualcomm Hexagon ~\cite{hexagon}.







A future direction we intend to pursue with the framework is the ability to perform cross layer optimization. Notice that all the layers have the same loop structure, and utilize the same data layout.
The common data format across layers lends itself particularly well to the identification and applicaiton of fusion since it simplifies fusion data dependence analysis. The concept of the abstract layer can be extended to be an abstract "fused" layer. This would allow a number of layers that fit the template to be fused without having to write a different implementation for every combination of layers.

We believe that the principles (common data layout and loop nest around performant kernels) underlying the SMaLL framework can support more complicated layers in other networks such as the attention layers in Transformers~\cite{bert, attention}. For example, an attention layer can be cast in terms of a scaled matrix multiplications (a specialized Full Layer) of the queries and keys followed by softmax (implemented as a Single Element Layer followed by either a Single Channel Layer for computing the sum of exponential values) and another matrix multiplication of the result with the values vectors (another Full layer). Individually, these layers can be described in terms of the specialized layer in SMaLL. However, in order for the implementation to be performant, fusing across multiple layers as discussed in the previous paragraph will be required.

\begin{acks}
This work was partially sponsored by the Defense Advanced Research Projects Agency (DARPA) TRIAD program under Agreement No. HR00112190099, and is based upon work funded and supported by the Department of Defense under Contract No. FA8702-15-D-0002 with Carnegie Mellon University for the operation of the Software Engineering Institute, a federally funded research and development center. 
This work is also partially sponsored by a gift from Meta.

The view, opinions, and/or findings contained in this material are those of the author(s) and should not be construed as an official Government position, policy, or decision, unless designated by other documentation.

References herein to any specific commercial product, process, or service by trade name, trade mark, manufacturer, or otherwise, does not necessarily constitute or imply its endorsement, recommendation, or favoring by Carnegie Mellon University or its Software Engineering Institute.
[DISTRIBUTION STATEMENT A] This material has been approved for public release and unlimited distribution.  Please see Copyright notice for non-US Government use and distribution. [DM22-1168]

\end{acks}
\bibliographystyle{unsrt}
\bibliography{fusion_ref}

\begin{thebibliography}{10}

\bibitem{matrix_engines}
Jos{\'{e}}~E. Moreira, Kit Barton, Steven Battle, Peter Bergner, Ramon Bertran,
  Puneeth Bhat, Pedro Caldeira, David Edelsohn, Gordon~C. Fossum, Brad Frey,
  Nemanja Ivanovic, Chip Kerchner, Vincent Lim, Shakti Kapoor, Tulio~Machado
  Filho, Silvia~Melitta Mueller, Brett Olsson, Satish Sadasivam, Baptiste
  Saleil, Bill Schmidt, Rajalakshmi Srinivasaraghavan, Shricharan Srivatsan,
  Brian~W. Thompto, Andreas Wagner, and Nelson Wu.
\newblock A matrix math facility for power {ISA(TM)} processors.
\newblock {\em CoRR}, abs/2104.03142, 2021.

\bibitem{a100}
Nvidia Corporation.
\newblock {\em {NVIDIA A100 Tensor Core GPU Architecture}}.

\bibitem{ml_chips}
Albert Reuther, Peter Michaleas, Michael Jones, Vijay Gadepally, Siddharth
  Samsi, and Jeremy Kepner.
\newblock Survey of machine learning accelerators.
\newblock {\em CoRR}, abs/2009.00993, 2020.

\bibitem{tensorflow}
Mart\'{i}n Abadi, Ashish Agarwal, Paul Barham, Eugene Brevdo, Zhifeng Chen,
  Craig Citro, Greg~S. Corrado, Andy Davis, Jeffrey Dean, Matthieu Devin,
  Sanjay Ghemawat, Ian Goodfellow, Andrew Harp, Geoffrey Irving, Michael Isard,
  Yangqing Jia, Rafal Jozefowicz, Lukasz Kaiser, Manjunath Kudlur, Josh
  Levenberg, Dandelion Man\'{e}, Rajat Monga, Sherry Moore, Derek Murray, Chris
  Olah, Mike Schuster, Jonathon Shlens, Benoit Steiner, Ilya Sutskever, Kunal
  Talwar, Paul Tucker, Vincent Vanhoucke, Vijay Vasudevan, Fernanda Vi\'{e}gas,
  Oriol Vinyals, Pete Warden, Martin Wattenberg, Martin Wicke, Yuan Yu, and
  Xiaoqiang Zheng.
\newblock {TensorFlow}: Large-scale machine learning on heterogeneous systems,
  2015.
\newblock Software available from tensorflow.org.

\bibitem{pytorch}
Adam Paszke, Sam Gross, Francisco Massa, Adam Lerer, James Bradbury, Gregory
  Chanan, Trevor Killeen, Zeming Lin, Natalia Gimelshein, Luca Antiga, Alban
  Desmaison, Andreas Kopf, Edward Yang, Zachary DeVito, Martin Raison, Alykhan
  Tejani, Sasank Chilamkurthy, Benoit Steiner, Lu~Fang, Junjie Bai, and Soumith
  Chintala.
\newblock Pytorch: An imperative style, high-performance deep learning library.
\newblock In {\em Advances in Neural Information Processing Systems 32}, pages
  8024--8035. Curran Associates, Inc., 2019.

\bibitem{onednn}
\text{Intel® oneAPI Deep Neural Network Library}.
\newblock Available at
  \url{https://www.intel.com/content/www/us/en/developer/tools/oneapi/onednn.html}
  (2022/10/31).

\bibitem{cmsis-nn}
Liangzhen Lai, Naveen Suda, and Vikas Chandra.
\newblock {CMSIS-NN:} efficient neural network kernels for arm cortex-m cpus.
\newblock {\em CoRR}, abs/1801.06601, 2018.

\bibitem{enrique_paper}
Sergio Barrachina, Adri{\'a}n Castell{\'o}, Manuel~F Dolz, Tze~Meng Low,
  H{\'e}ctor Mart{\'\i}nez, Enrique~S Quintana-Ort{\'\i}, Upasana Sridhar, and
  Andr{\'e}s~E Tom{\'a}s.
\newblock Reformulating the direct convolution for high-performance deep
  learning inference on arm processors.
\newblock {\em Journal of Systems Architecture}, 135:102806, 2023.

\bibitem{im2col}
Kumar Chellapilla, Sidd Puri, and Patrice Simard.
\newblock {High Performance Convolutional Neural Networks for Document
  Processing}.
\newblock In Guy Lorette, editor, {\em {Tenth International Workshop on
  Frontiers in Handwriting Recognition}}, La Baule (France), October 2006.
  {Universit{\'e} de Rennes 1}, {Suvisoft}.
\newblock http://www.suvisoft.com.

\bibitem{TVM}
Tianqi Chen, Thierry Moreau, Ziheng Jiang, Lianmin Zheng, Eddie Yan, Meghan
  Cowan, Haichen Shen, Leyuan Wang, Yuwei Hu, Luis Ceze, Carlos Guestrin, and
  Arvind Krishnamurthy.
\newblock Tvm: An automated end-to-end optimizing compiler for deep learning,
  2018.

\bibitem{XLA}
Chris Leary and Todd Wang.
\newblock Xla: Tensorflow, compiled.
\newblock 2017.

\bibitem{glow}
Nadav Rotem, Jordan Fix, Saleem Abdulrasool, et~al.
\newblock Glow: Graph lowering compiler techniques for neural networks.
\newblock {\em arXiv preprint arXiv:1805.00907}, 2018.

\bibitem{BLIS1}
Field~G. Van~Zee and Robert~A. van~de Geijn.
\newblock Blis: A framework for rapidly instantiating blas functionality.
\newblock {\em ACM Trans. Math. Softw.}, 41(3), jun 2015.

\bibitem{BLAS1}
C.~L. Lawson, R.~J. Hanson, F.~T. Krogh, and D.~R. Kincaid.
\newblock Algorithm 539: Basic linear algebra subprograms for fortran usage
  [f1].
\newblock {\em ACM Trans. Math. Softw.}, 5(3):324–325, sep 1979.

\bibitem{BLAS2}
Jack~J. Dongarra, Jeremy Du~Croz, Sven Hammarling, and Richard~J. Hanson.
\newblock An extended set of fortran basic linear algebra subprograms.
\newblock {\em ACM Trans. Math. Softw.}, 14(1):1–17, mar 1988.

\bibitem{BLAS3}
Jack~J. Dongarra, Jeremy Du~Croz, Sven Hammarling, and Iain Duff.
\newblock A set of level 3 basic linear algebra subprograms.
\newblock {\em ACM Transactions on Mathematical Software}, 16(1):1--17, March
  1990.

\bibitem{goto}
Kazushige Goto and Robert A. Van~De Geijn.
\newblock Anatomy of high-performance matrix multiplication.
\newblock {\em ACM Transactions on Mathematical Software}, pages 1--25, 2008.

\bibitem{BLIS2}
Field~G. {V}an {Z}ee, Tyler Smith, Francisco~D. Igual, Mikhail Smelyanskiy,
  Xianyi Zhang, Michael Kistler, Vernon Austel, John Gunnels, Tze~Meng Low,
  Bryan Marker, Lee Killough, and Robert~A. {v}an~{d}e {G}eijn.
\newblock The {BLIS} framework: Experiments in portability.
\newblock {\em ACM Transactions on Mathematical Software}, 42(2):12:1--12:19,
  June 2016.

\bibitem{blis4}
Tze~Meng Low, Francisco~D. Igual, Tyler~M. Smith, and Enrique~S.
  Quintana-Ort\'{\i}.
\newblock Analytical modeling is enough for high-performance {BLIS}.
\newblock {\em ACM Transactions on Mathematical Software}, 43(2):12:1--12:18,
  August 2016.

\bibitem{VGGnet}
Karen Simonyan and Andrew Zisserman.
\newblock Very deep convolutional networks for large-scale image recognition.
\newblock In {\em International Conference on Learning Representations}, 2015.

\bibitem{resnet}
Kaiming He, Xiangyu Zhang, Shaoqing Ren, and Jian Sun.
\newblock Deep residual learning for image recognition.
\newblock {\em CoRR}, abs/1512.03385, 2015.

\bibitem{Inception}
Christian Szegedy, Wei Liu, Yangqing Jia, Pierre Sermanet, Scott Reed, Dragomir
  Anguelov, Dumitru Erhan, Vincent Vanhoucke, and Andrew Rabinovich.
\newblock Going deeper with convolutions.
\newblock In {\em 2015 IEEE Conference on Computer Vision and Pattern
  Recognition (CVPR)}, pages 1--9, 2015.

\bibitem{direct_conv}
Jiyuan Zhang, Franz Franchetti, and Tze~Meng Low.
\newblock High performance zero-memory overhead direct convolutions.
\newblock In {\em International Conference on Machine Learning}, pages
  5776--5785. PMLR, 2018.

\bibitem{BLIS4x}
Tze~Meng Low, Francisco~D. Igual, Tyler~M. Smith, and Enrique~S.
  Quintana-Ort\'{\i}.
\newblock Analytical modeling is enough for high-performance {BLIS}.
\newblock {\em ACM Transactions on Mathematical Software}, 43(2):12:1--12:18,
  August 2016.

\bibitem{arm-cortex}
{Robert Triggs}.
\newblock {Qualcomm Snapdragon 888 deep dive: Everything you need to know}.

\bibitem{Snapdragon}
Qualcomm Corporation.
\newblock {\em {Snapdragon 888 Specifications}}.

\bibitem{intel_extensions}
\text{Intel® Intrinsics Guide}.
\newblock Available at
  \url{https://www.intel.com/content/www/us/en/docs/intrinsics-guide/index.html}
  (2022/10/31).

\bibitem{autoTVM}
Tianqi Chen, Lianmin Zheng, Eddie Yan, Ziheng Jiang, Thierry Moreau, Luis Ceze,
  Carlos Guestrin, and Arvind Krishnamurthy.
\newblock Learning to optimize tensor programs.
\newblock {\em Advances in Neural Information Processing Systems}, 31, 2018.

\bibitem{tflite}
\text{TensorFlow Lite | ML for Mobile and Edge Devices}.
\newblock Available at \url{https://www.tensorflow.org/lite/} (2022/02/04).

\bibitem{SNPE}
Qualcomm Corporation.
\newblock {\em {Snapdragon Neural Processing Engine SDK}}.

\bibitem{arduino-ide}
Arduino.
\newblock {\em {Arduino IDE}}.

\bibitem{arduino-cli}
Arduino.
\newblock {\em {Arduino Command-Line Interface}}.

\bibitem{mlcommons_tiny}
Colby Banbury, Vijay~Janapa Reddi, Peter Torelli, Jeremy Holleman, Nat
  Jeffries, Csaba Kiraly, Pietro Montino, David Kanter, Sebastian Ahmed, Danilo
  Pau, et~al.
\newblock Mlperf tiny benchmark.
\newblock {\em Proceedings of the Neural Information Processing Systems Track
  on Datasets and Benchmarks}, 2021.

\bibitem{ds-conv}
L~Sifre and St{\'e}phane Mallat.
\newblock Rigid-motion scattering for texture classification [j].
\newblock {\em Computer Science}, 3559:501--515, 2014.

\bibitem{mobileNetv1}
Andrew~G. Howard, Menglong Zhu, Bo~Chen, Dmitry Kalenichenko, Weijun Wang,
  Tobias Weyand, Marco Andreetto, and Hartwig Adam.
\newblock Mobilenets: Efficient convolutional neural networks for mobile vision
  applications.
\newblock {\em CoRR}, abs/1704.04861, 2017.

\bibitem{DS-CNN}
Peter~M{\o}lgaard S{\o}rensen, Bastian Epp, and Tobias May.
\newblock A depthwise separable convolutional neural network for keyword
  spotting on an embedded system.
\newblock {\em EURASIP Journal on Audio, Speech, and Music Processing},
  2020(1):1--14, 2020.

\bibitem{keras}
\text{Keras}.
\newblock Available at \url{https://keras.io/} (2022/10/31).

\bibitem{tf-lite-bench}
Tensorflow.
\newblock {\em {Performance measurement with TF-Lite}}.

\bibitem{monica_lam}
Monica Lam.
\newblock Software pipelining: An effective scheduling technique for vliw
  machines.
\newblock In {\em Proceedings of the ACM SIGPLAN 1988 conference on Programming
  Language design and Implementation}, pages 318--328, 1988.

\bibitem{jetson}
\text{Advanced AI Embedded Systems}.
\newblock Available at
  \url{https://www.nvidia.com/en-us/autonomous-machines/embedded-systems/}
  (2022/10/31).

\bibitem{hexagon}
\text{Hexagon DSP Processor}.
\newblock Available at
  \url{https://developer.qualcomm.com/software/hexagon-dsp-sdk/dsp-processor}
  (2022/10/31).

\bibitem{bert}
Jacob Devlin, Ming-Wei Chang, Kenton Lee, and Kristina Toutanova.
\newblock Bert: Pre-training of deep bidirectional transformers for language
  understanding.
\newblock {\em arXiv preprint arXiv:1810.04805}, 2018.

\bibitem{attention}
Ashish Vaswani, Noam Shazeer, Niki Parmar, Jakob Uszkoreit, Llion Jones,
  Aidan~N Gomez, {\L}ukasz Kaiser, and Illia Polosukhin.
\newblock Attention is all you need.
\newblock {\em Advances in neural information processing systems}, 30, 2017.

\end{thebibliography}
\end{document}